\documentstyle[11pt]{article}
\begin{document}
\title{\Large \bf Rotation of D-brane and Non-commutative Geometry}
\author
{Pei Wang\footnote{peiwang@nwu.edu.cn}\hspace{1cm} and \hspace{1cm} 
Ruihong
Yue\footnote{yue@phy.nwu.edu.cn}\\
 Institute of Modern Physics, Northwest University, P.O.Box 105,\\
Xi'an 710069, P.R. China}
\maketitle
%\begin{abstract}
\begin{center}{Abstract}
\end{center}
Our motivation is to find the relationship between the commutator 
of coordinates and uncertainty relation involving only the coordinates.
The boundary condition with constant background field is connected with 
the rotation of D-brane at general angle. And the mode expansions of
D-brane we found is more reasonable than those appeared in literature. 
The partition functions and scattering amplitudes are also discussed.



\section{Introduction}
~
     
The non-commutative geometry is one of the most important and most 
interesting idea in the recent string theory. [1] In an excellent paper 
Seiberg and Witten express the commutator of coordinates in terms of a
constant Neveu-Schwarz B-field (see [2]
 and references therein). On the other hand an uncertainty relation 
involving only the coordinates was appeared in the Polchinski's wonderful 
book,[3] which is developed from some discussions about dynamics of 
D-branes.[4,5,6] As we know from ordinary quantum mechanics the
uncertainty relation and the commutator between coordinate and 
momentum both involve a Planck constant (over $2\pi$  )$\hbar$ 
 and henceforth are related each other. Naturally, we are interested in
knowing if there is any relationship between the commutator of 
coordinates and the uncertainty relation in the coordinates 
in the domain of  non-commutative geometry.

We will find in this paper that the boundary conditions with 
constant background B-field, which is in a special form, 
are equivalent to the rotation of D-branes 
(including their analytic continuation, the relative motion of D-branes).
So we may use these properties of D-brane to seek the 
relationship mentioned above.

The rotations of D-branes are discussed by several authors.[3,7]
However, it seems to us that the mode expansions are hard to reduce into
the parallel D-branes (including NN, DD, ND, DN coordinates) when the 
rotated angles vanish (or $\pi/2$ , in case where vanishing  angle leads
to perpendicular D-branes ). By using the above equivalence we find new
mode expansions proposed in this paper is more convincing. That is also
the main part of our positive results.

Given the relevant mode expansions we can calculate the partition 
functions easily. In concrete, we study D0-D4 and D4-D4 open strings. They
have different partition functions but same non-vanishing  potential. 
By means of analytic continuation we find the  scattering amplitudes 
of relative motion of D-branes identical with one given by Polchinski.
But it is difficult to  interpret consistently  the relationship
between commutator of pure coordinates and uncertainty relation 
in small separation.   

\section{Non-commutative Geometry and the Rotation of D-brane}
~

 Following Seiberg and Witten [2] the commutators of coordinates 
can be written as
\begin{equation}
        \left[ X^i, X^j  \right ]= i \theta^{ij}
\end{equation}              
Under a nonzero constant background the $\theta$  can be expressed 
in terms of B-field. Especially in the zero slope limit 
( $\alpha'\sim \epsilon^{1/2}\rightarrow 0,\,
g_{ij}\sim\epsilon\rightarrow 0$)
we have $\theta=1/B$, but we will not restrict
ourselves in this limit. In this paper, we set
$g_{ij}=\epsilon\delta_{ij}$, and B in a special form
\begin{equation}
B=\frac{\epsilon}{2\pi\alpha'}\left(\begin{array}{cccc}
  0 & b_1 & 0 & 0 \\-b_1 & 0 & 0& 0 \\ 0 &0 & 0& b_2\\ 0&0&-b_2& 0
\end{array} \right)
\end{equation} 
       
     Now, let us consider an open string stretched between two D-branes.
The first example is D0-D4 open string which has following boundary
conditions ( $i,j=1,\cdots,4$ )
\begin{equation} 
\partial_{\sigma} X^0|_{\sigma=0,\pi}=\partial_{\tau}
X'^{5\cdots 9}|_{\sigma=0,\pi}=\partial_{\tau} X^{1\cdots4}|_{\sigma=0}=
g_{ij}\partial_{\sigma} X^j+2\pi\alpha'B_{ij}\partial_{\tau}
X^j|_{\sigma=\pi}=0
\end{equation}
in which we write $X'$  for dimensions $5\cdots 9$, 
because of the fact that from the last boundary condition the D4 brane is
not perpendicular to the open string directions at $\sigma=\pi$ ( it
represents that the brane has sloped or deformed under interaction with
background field ), so in general, $X'^{5\cdots 9}$
may not  orthogonalize $X^{1\cdots 4}$.  To get a mode expansions
for$X^{1\cdots 4}$, Seiberg and
Witten introduced following "complex" scalar fields:
$Z_1=X_1+iX_2,\,Z_2=X_3+iX_4$ and
$\bar{Z}_1=X_1-iX_2,\,\bar{Z}_2=X_3-iX_4$. 
But $\bar{Z}_{I=1,2}$ are not the complex conjugate fields of
${Z}_{I=1,2}$, because that $X_{1\cdots 4}$ can
have their own complex mode expansions unless $X_{1\cdots 4}$ are
hermitian, and then $b_{1,2}$ 
are pure imaginary. In this regime the boundary condition at $\sigma=\pi$ 
changes to [2]  
\begin{equation}
\partial_{\sigma}Z_I+b_I\partial_{\tau}Z_I=0\quad, \quad
\partial_{\sigma}\overline{Z}_I-b_I\partial_{\tau}\overline{Z}_I=0\, .
\end{equation} 
Latter we refer it the twisted condition. 
The mode expansions have been expressed by them as following
\begin{eqnarray}
Z_I&=&i\sqrt{\frac{\alpha'}{2}}\sum_n
     \left(e^{(n+\nu_I)(\tau+i\sigma)}-e^{(n+\nu_I)(\tau-i\sigma)}\right)
     \frac{\alpha^I_{n+\nu_I}}{n+\nu_I}\nonumber \\
\overline{Z}_I&=&i\sqrt{\frac{\alpha'}{2}}\sum_n
     \left(e^{(n-\nu_I)(\tau+i\sigma)}-e^{(n-\nu_I)(\tau-i\sigma)}\right)
     \frac{\overline{\alpha}^I_{-n+\nu_I}}{n-\nu_I}
\end{eqnarray}
They satisfy  
\begin{equation}
\partial_{\tau}Z_I|_{\sigma=0}=
\partial_{\tau}\overline{Z}_I|_{\sigma=0}=0
\end{equation}
and                                                               
\begin{eqnarray}
b_I&=&-\frac{\partial_{\sigma}Z_I}{\partial_{\tau}Z_I}|_{\sigma=\pi}=
\frac{\partial_{\sigma}\overline{Z}_I}{\partial_{\tau}
\overline{Z}_I}|_{\sigma=\pi}
\nonumber \\
 &=&-i\frac{e^{i\pi\nu_I}+e^{-i\pi\nu_I}}{e^{i\pi\nu_I}-e^{-i\pi\nu_I}}
 =-\cot(\pi\nu_I)\equiv-\cot(\varphi_I)
\end{eqnarray}
In an alternative description we can rewrite the boundary 
conditions (4) as follows
\begin{eqnarray}                                
\left(\cos(\varphi_I)\partial_{\tau}-\sin(\varphi_I)\partial_{\sigma}\right)
Z_I|_{\sigma=\pi} &=& \partial_{\sigma'}Z_I|_{\sigma=\pi}=0\nonumber \\
\left(\cos(\varphi_I)\partial_{\tau}+\sin(\varphi_I)\partial_{\sigma}\right)
\overline{Z}_I|_{\sigma=\pi} &=& \partial_{\sigma"}
\overline{Z}_I|_{\sigma=\pi}=0
\end{eqnarray}
in which we denote new world sheet coordinates for
reference, which are
in parallel and vertical directions of the D4 brane
\begin{eqnarray}
\sigma'&=&\cos(\varphi_I)\tau-\sin(\varphi_I)\sigma \quad,\quad
\tau'=\sin(\varphi_I)\tau+\cos(\varphi_I)\sigma\quad,\nonumber \\
\sigma"&=&\cos(\varphi_I)\tau+\sin(\varphi_I)\sigma \quad,\quad
\tau"=-\sin(\varphi_I)\tau+\cos(\varphi_I)\sigma\quad.
\end{eqnarray}
   
  To obtain the partition function of full string we should consider 
the contributions from another independent dimensions which are
orthogonalized with above $Z_I$  and $\overline{Z}_I$, but not from 
$X'_{5\cdots 9}$. To this end let us define                                    
\begin{equation}
Z_M=X_{2M-1}+iX_{2M}\quad,\quad \overline{Z}_M=X_{2M-1}-iX_{2M}\quad,\quad
(M=I;J=1,2; 3,4)
\end{equation}
Except $Z_I$ and $\overline{Z}_I$, we introduce their
counterparts  $Z_J$ and $\overline{Z}_J$ respectively. 
Hence they observe boundary conditions 
\begin{equation}
\partial_{\tau}Z_J|_{\sigma=0}=\partial_{\tau}
\overline{Z}_J|_{\sigma=0}=0 
\end{equation}
and                                                                                  
\begin{eqnarray}
\left(\sin(\varphi_I)\partial_{\tau}+\cos(\varphi_I)\partial_{\sigma}\right)
Z_J|_{\sigma=\pi} &=& \partial_{\tau'}Z_J|_{\sigma=\pi}=0,\nonumber \\
\left(-\sin(\varphi_I)\partial_{\tau}+\cos(\varphi_I)\partial_{\sigma}\right)
\overline{Z}_J|_{\sigma=\pi} &=& \partial_{\tau"}
\overline{Z}_J|_{\sigma=\pi}=0.
\end{eqnarray}
Therefore we  find their mode expansions easily
\begin{eqnarray}
Z_J&=&i\sqrt{\frac{\alpha'}{2}}\sum_n
\left(e^{(n+\nu_I+1/2)\omega}
      -e^{(n+\nu_I+1/2)\overline{\omega}}\right)
     \frac{\alpha^I_{n+\nu_I+1/2}}{n+\nu_I+1/2}\nonumber \\
\overline{Z}_I&=&i\sqrt{\frac{\alpha'}{2}}\sum_n
\left(e^{(n-\nu_I-1/2)\omega}
-e^{(n-\nu_I-1/2)\overline{\omega}}\right)
     \frac{\overline{\alpha}^I_{-n+\nu_I+1/2}}{n-\nu_I-1/2}
\end{eqnarray}                   
where $\omega=\sigma+i\tau, \overline{\omega}=\sigma-i\tau$                
It is very interesting that the boundary conditions (8) and (12) are
equivalent to the rotation of D4 brane                         
\begin{eqnarray}
\partial_{\sigma}\left(-\sin(\varphi_I)Z_I
+\cos(\varphi_I)Z_J\right)|_{\sigma=\pi}&\equiv&
\partial_{\sigma}Z'_I|_{\sigma=\pi}=0, \nonumber \\
\partial_{\sigma}\left(\sin(\varphi_I)\overline{Z}_I
+\cos(\varphi_I)\overline{Z}_J\right)|_{\sigma=\pi}&\equiv&
\partial_{\sigma}\overline{Z}"_I|_{\sigma=\pi}=0, \nonumber \\
\partial_{\tau}\left(\cos(\varphi_I)Z_I
+\sin(\varphi_I)Z_J\right)|_{\sigma=\pi}&\equiv&
\partial_{\tau}Z'_I|_{\sigma=\pi}=0, \nonumber \\
\partial_{\tau}\left(\cos(\varphi_I)\overline{Z}_I
-\sin(\varphi_I)\overline{Z}_J\right)|_{\sigma=\pi}&\equiv&
\partial_{\tau}\overline{Z}"_I|_{\sigma=\pi}=0.
\end{eqnarray}
In fact if we assume that                                        
$$
Z_I=-SZ'_I+CZ'_J\quad,\quad Z_J=CZ'_I+SZ'_J 
$$
in which
$$
\partial_{\sigma}Z'_I=\partial_{\tau}Z'_J=0
$$
From Eq. (8) and Eq. (12),  we have                                             
$$
\cos(\varphi_I)S\partial_{\tau}Z'_I
+\sin(\varphi_I)C\partial_{\sigma}Z'_J=0
$$
and                                                                       
$$
\sin(\varphi_I)C\partial_{\tau}Z'_I+\cos(\varphi_I)S\partial_{\sigma}Z'_J=0
$$                                                                  
Eliminating  $\partial_{\tau}Z'_I$  from
these two equations we obtain
$$
\cos^2(\varphi_I) S^2\partial_{\sigma}Z'_J=\sin^2(\varphi_I)
C^2\partial_{\sigma}Z'_J.
$$
The solution of this equation is                                   
$$
S=\sin(\varphi_I)\quad,\quad C=\cos(\varphi_I) 
$$      
In a similar method we can show that                                  
$$
\overline{Z}"_I=\sin(\varphi_I)\overline{Z}_I
 +\cos(\varphi_I)\overline{Z}_J\quad,\quad
\overline{Z}"_J=\cos(\varphi_I)\overline{Z}_I-\sin(\varphi_I)
\overline{Z}_J
$$                                                               
Like Eq.(10)  we suppose that $Z'_I,\overline{Z}"_I,Z'_J,\overline{Z}"_J$
 are combinations from orthogonal systems $X'_m$ and  $X"_m$
($m=0,1,\cdots,9$) with $ X'_{0,9}=X"_{0,9}=X_{0,9}$.       
That is to say systems $X'_m$ and $X"_m$ are different from $ X_m$ by a
relative angle around $ X_9$.

\section{D4-D4 Open String} 
     Similarly we can also consider two D4 branes with relative rotation
angles. The boundary conditions for this system are                    
\begin{equation}
\partial_{\sigma}Z_I|_{\sigma=0}
=\partial_{\sigma}\overline{Z}_I|_{\sigma=0}
=\partial_{\tau}Z_J|_{\sigma=0}
=\partial_{\tau}\overline{Z}_J|_{\sigma=0}=0
\end{equation}
and conditions (14) at $\sigma=\pi$. Thus, the mode
expansions of $Z_J$ and $\overline{Z}_J$   are still given by Eq.(13).  
As for $Z_I$ and $\overline{Z}_I$ we have following expressions       
\begin{eqnarray}
Z_I&=&i\sqrt{\frac{\alpha'}{2}}\sum_n\left(e^{(n+\nu_I+1/2)\omega}
     +e^{(n+\nu_I+1/2)\overline{\omega}}\right)
         \frac{\alpha^I_{n+\nu_I+1/2}}{n+\nu_I+1/2},\nonumber \\
\overline{Z}_I&=&i\sqrt{\frac{\alpha'}{2}}\sum_n\left(e^{(n-\nu_I-1/2)\omega}
     +e^{(n-\nu_I-1/2)\overline{\omega}}\right)
         \frac{\overline{\alpha}^I_{-n+\nu_I+1/2}}{n-\nu_I-1/2}
\end{eqnarray}
Here we find same exponentials in Eq.(13) and Eq.(16), so it is convenient
to use 
\begin{equation}
\tilde{\nu}_I=\nu_I+1/2\quad \left(\tilde{\varphi}_I=\varphi_I+\pi/2
=\tan^{-1}(b_I) \right)
\end{equation}                                           
instead of  $\nu_I$ for D4-D4 string.

\section{ Branes at general angles}
~                           

Till now we have shown that the twisted boundary conditions are equivalent 
to a rotation of D-brane: an angle $\varphi_1$  in the (1,5) and (2,6)
planes;  $\varphi_2$ in
the (3,7) and (4,8 ) planes. As a matter of fact, we can still 
generalize it to more general case, that is a rotation with different 
angles in (1,5) and (2,6) planes and different angles in (3,7) and 
(4,8) planes. To show the generalized mode expansion we need only give an
illustration: A rotation of angle $\varphi$ in one plane, say (1,5). We
will give up to use the $Z_M$  fields, but utilize the original $X_m$ 
field. Reversing the above mentioned trick the boundary conditions
\begin{equation}
\partial_{\sigma}\left(-\sin(\varphi)X_1+\cos(\varphi)X_5\right)
|_{\sigma=\pi}=
\partial_{\tau}\left(\cos(\varphi)X_1+\sin(\varphi)X_5\right)
|_{\sigma=\pi}=0
\end{equation}
are equivalent to the twisted one                                 
\begin{equation}
\left(\cos(\varphi)\partial_{\tau}-\sin(\varphi)\partial_{\sigma}\right)
X_1|_{\sigma=\pi}=
\left(\sin(\varphi)\partial_{\tau}+\cos(\varphi)\partial_{\sigma}\right)
X_5|_{\sigma=\pi}=0.
\end{equation} 
Therefore the mode expansions of $X_1$   and $X_5$   are written as
follows
\begin{eqnarray}
X_1&=&i\sqrt{\frac{\alpha'}{2}}\sum_n\left(e^{(n+\nu)\omega}
     -e^{(n+\nu)\overline{\omega}}\right)
         \frac{\alpha^I_{n+\nu}}{n+\nu}\quad,\quad
\mbox{for}\quad D0-D4, \nonumber \\
X_1&=&i\sqrt{\frac{\alpha'}{2}}\sum_n\left(e^{(n+\tilde{\nu})\omega}
     +e^{(n+\tilde{\nu})\overline{\omega}}\right)
\frac{\overline{\alpha}^I_{n+\tilde{\nu}}}{n+\tilde{\nu}}
\quad,\quad \mbox{for} \quad D4-D4.\\
%\end{eqnarray}
%\begin{eqnarray}
X_5&=&i\sqrt{\frac{\alpha'}{2}}\sum_n\left(
e^{(n+\nu+1/2)\omega}-e^{(n+\nu+1/2)\bar\omega}\right)
    \frac{\alpha_{n+\nu+1/2}}{n+\nu+1/2}\quad, \quad
\mbox{for}\quad D0-D4\nonumber  \\
&=&i\sqrt{\frac{\alpha'}{2}}\sum_n\left(
     e^{(n+\tilde\nu)\omega}-e^{(n+\tilde\nu)\bar\omega}\right)
     \frac{\alpha_{n+\tilde\nu}}{n+\tilde\nu}\quad, \quad
\mbox{for} \quad D4-D4
\end{eqnarray}
     Obviously these expressions include the parallel branes as  their
special cases. Eq.(20a) reduces to DD open string when $\nu=0$, DN
when $\nu=1/2$; and the case of Eq.(21) is just the opposite.
Nevertheless, eq.(20b) will give NN string when $\tilde\nu=0$ and ND 
when $\tilde\nu=1/2$   .
\section{ World sheet Supersymmetry} 
     Let us go back to the issue in which there are given  background
fields B. The same conclusion about the equivalence between  the twisted
boundary conditions and rotation of D-brane can find from the  boundary
conditions of world sheet fermions. The latter are [2]
\begin{eqnarray}
\overline\psi_i&=&R_{ij}(B)\psi^j  \nonumber \\
R(B)&=&(1-2\pi\alpha'g^{-1}B)^{-1}(1+2\pi\alpha'g^{-1}B) 
\nonumber \\[3mm]
    &=&\left( \begin{array}{cccc}
       -\cos(2\varphi_1) & -\sin(2\varphi_1) & 0 & 0\\
        \sin(2\varphi_1) & -\cos(2\varphi_1) & 0 & 0\\
        0 & 0 & -\cos(2\varphi_2) & -\sin(2\varphi_2)\\
        0 & 0 & \sin2\varphi_2) & -\cos(2\varphi_2)\end{array}
  \right )
\end{eqnarray}
One may consider a rotation so that                              
\begin{eqnarray}
 \overline\Psi_i&=&(\rho^{-1}(B))_{ij}\overline{\psi}^j\quad, \quad
  \Psi_i=\rho_{ij}(B)\psi^j \quad,\nonumber \\
\rho(B)&=&\left( \begin{array}{cccc}
            \sin(\phi_1) & -\cos(\phi_1) & 0 & 0\\
            \cos(\phi_1) & \sin(\phi_1) & 0 & 0\\
            0 & 0 & \sin(\phi_2) &  -\cos(\phi_2)\\
            0 & 0 & \cos(\phi_2) & \sin(\phi_2) \end{array}\right ) =
 (R(B))^{\frac{1}{2}}.
\end{eqnarray}

Then, after the rotation $\rho(B)$ the boundary condition will become a
simple form                                                      
\begin{equation}
 \overline\Psi_i=\Psi_i 
\end{equation}                                       
Once-more, we  have proven that the boundary condition with constant
background field is equivalent to a rotation of D-brane.
Rotation matrix    is a special subgroup of U(4),[3]
 \begin{equation} 
diag(exp(i\phi_1),exp(i\phi_2),exp(i\phi_3),exp(i\phi_4)) 
\end{equation}
when $\phi_1=\phi_2=\varphi_1,\;\;\phi_3=\phi_4=\varphi_2$. As we know a
D4
brane preserves super-charges of the form $\epsilon_L Q^L + \epsilon_R Q^R
$ with                     
 \begin{equation}
\epsilon_R = \Gamma_0\Gamma_1\cdots\Gamma_4\epsilon_L =
\beta^\perp \epsilon_L,
\end{equation}  
in which Polchinski's symbol $\beta^\perp$    is also used. The
supersymmetry unbroken by the rotation 4-brane requires 
\begin{equation}
\rho^{-1}\beta^\perp \rho = \beta^\perp \rho^2 = \beta^\perp,
\end{equation}                   
and                                                          
\begin{equation}
\rho^2 = \exp(2 i \sum_{a=1}^4 s_a \phi_a) 
\end{equation}  
is the rotation in the usual spinor basis.[3] Since $\phi_1 = \phi_2,\;\; 
\phi_3 =
\phi_4$ 
in our case, there are 1/4 unbroken supersymmetries of total 16, if
$\phi_a\neq 0$.
    
Although there are unbroken supersymmetries the potential is nonzero.

\section{ Partition Functions }
From mode expansions (5),(13),(16) we know                           
\begin{equation}
[ \alpha_{n+\nu},\overline{\alpha}_{m+\nu}]=(n+\nu)\delta_{n,-m}\quad,
\quad\nu=\nu_I \;\;or\;\; \tilde{\nu_I} 
\end{equation}
It is easy to find the partition function for one such "complex" scalar
(see for example[3])                             
\begin{equation}
-i\frac{exp(\pi \nu^{2}t)\eta(it)}{\theta_{11}(i\nu t,it)}
\end{equation}
On account of [8]                                                
\begin{eqnarray}
Z_{\beta}^{\alpha}(\pi(\nu +\frac{1}{2}),it)&=&\frac{\theta_{\alpha \beta}
(i(\nu +\frac{1}{2})t,it)}{exp(\pi(\nu
+\frac{1}{2})^{2}t)\eta(it)}\nonumber\\
&=&\frac{\theta_{\bar{\alpha}\bar{\beta}}
(i\nu
t,it)}{exp(\pi\nu^{2}t)\eta(it)}\equiv 
Z_{\bar{\beta}}^{\bar{\alpha}}(\pi\nu ,it) 
\end{eqnarray} 
The correspondence between two index pairs is
$$                                
\begin{tabular}{l|cccc}
$\alpha\beta$ & 00 & 01& 10&11\\ \hline
$\bar{\alpha}\bar{\beta}$& 01& 00 & 11 & 10
\end{tabular} 
$$
The full contribution of scalars is                                  
\begin{equation}
\prod_{I=1,2}(Z_{1}^{1}(\pi\nu_{I},it)Z^{1}_{0}(\pi\nu_{I},it))^{-1}
\quad,\quad \mbox{for}\quad D0-D4   
\end{equation}
and                                                          
\begin{equation}
\prod_{I=1,2}(Z_{1}^{1}(\pi \tilde{\nu_{I}},it))^{-2}\quad,\quad
\mbox{for}\quad D4-D4
\end{equation}  
And the related fermionic partition function is                       
\begin{eqnarray}
& &\frac{1}{2}[\prod_{I=1,2}Z_{0}^{0}(\pi\nu_{I},it)
   Z_{1}^{0}(\pi\nu_{I},it) 
  -Z_{1}^{0}(\pi\nu_{I},it)Z_{0}^{0}(\pi\nu_{I},it)\nonumber \\
& &-\prod_{I=1,2}Z_{0}^{1}(\pi\nu_{I},it)Z_{1}^{1}(\pi\nu_{I},it)
-\prod_{I=1,2}Z_{1}^{1}(\pi\nu_{I},it)Z_{0}^{1}(\pi\nu_{I},it)]
                    \nonumber\\
&=&\prod_{I=1,2}Z_{0}^{1}(\pi\nu_{I},it)Z_{1}^{1}(\pi\nu_{I},it)
\quad,\quad \mbox{for} \quad D0-D4 
\end{eqnarray}
and             
\begin{eqnarray}
& &\frac{1}{2}\left[\prod_{I=1,2}\left(
Z_{0}^{0}(\pi\tilde{\nu}_{I},it)\right)^2-\prod_{I=1,2}\left(
Z_{1}^{0}(\pi\tilde{\nu}_{I},it)\right)^2-\prod_{I=1,2}\left(
Z_{0}^{1}(\pi\tilde{\nu}_{I},it)\right)^2-\prod_{I=1,2}\left(
Z_{1}^{1}(\pi\tilde{\nu}_{I},it)\right)^2\right] \nonumber\\          
&=&\prod_{a=1}^4\left( Z_{1}^{1}(\pi\tilde{\nu}'_{a},it)\right)^2
=-\prod_{I=1,2}\left( Z_{1}^{1}(\pi\tilde{\nu}_{I},it)\right)^2
   \quad   \mbox{for}\quad D4-D4 
\end{eqnarray}
in which                
\begin{eqnarray} 
\nu'_1&=&\frac{1}{2}(\nu_1+\nu_1+\nu_2-\nu_2)=\nu_1 \nonumber\\
\nu'_2&=&\frac{1}{2}(\nu_1+\nu_1-\nu_2+\nu_2)=\nu_1 \nonumber\\
\nu'_3&=&\frac{1}{2}(\nu_1-\nu_1+\nu_2+\nu_2)=\nu_2 \nonumber\\
\nu'_4&=&\frac{1}{2}(\nu_1-\nu_1-\nu_2-\nu_2)=-\nu_2 
\end{eqnarray}                                     
from Riemann formula, and hence we have the last equality in (35) .
For the generalization  to general rotated angle, the results are the
same as ref.[3]\footnote{   Because of the last term in relevant Riemann
formula
is positive [8], we must change sign for one of $\phi_a$  in (13.4.22) in
reference[3].}

\section{Discussion for Dynamics} 

     In this section, let us focus on the D4-D4 case. By  imitating an
analysis made by Polchinski we may study the interaction of  D-branes by
means of analytic continuation. Since any $X^i$ on D4 brane has
Neumann boundary condition just like the "time" component $X^0$.
 We can analytically continue one of $X^i$ to be the time, say
$X^1\rightarrow i\tilde{X}^0$ ( $X^0$ continues to a space dimension ).
Then the rotation in (1,5) plane will become  a relative motion with
velocity 
\begin{equation}
v = i b_1 = i\tan(\tilde{\varphi}_1) = \tanh(\tilde{u}), \hspace{1cm}
\tilde{u} = i \tilde{\varphi}.
\end{equation}
Now, the full answer is a combination of a rotation
$(0, \tilde{\varphi}_1, \tilde{\varphi}_2, \tilde{\varphi}_2)$,
and a relative motion $(-i\tilde{u}, 0, 0, 0)$. According to ref[3]
the scattering amplitude is expressed as follows  
\begin{eqnarray}
\cal{A} & = & -i\int_{-\infty}^\infty d\tau V(r(\tau), v),\hspace{1cm}
r(\tau)^2 = y^2 + v^2\tau^2.\\\nonumber
V(r, v) & = &
i\frac{2V_4}{(8\pi^2\alpha^\prime)^{\frac{5}{2}}}\int_0^\infty
dt~t^{1/2}\exp(-\frac{t r^2}{2\pi\alpha^\prime})\frac{\tanh
(\tilde{u})\theta_{11}(\frac{i\tilde{u}}{2\pi}, \frac{i}{t})^4}{\eta(i
t)^9\theta_{11}(\frac{i\tilde{u}}{\pi},\frac{i}{t})}
\end{eqnarray}
Where $y$ is a separation  along the 9-direction (here we use the same
$\tau$ to denote Minkowski time, not  the Euclidean one). Polchinski
analyzed the
small $v$ ( so $v\approx \tilde{u}$ ) and even small $r$ case for 
nontrivial potential (see also [6]). That is  $ut \approx
\frac{\alpha^\prime v}{r_2} \approx 1$. Let
$\delta x \approx r, \delta t \approx \frac{r}{v}$, he
wrote down an uncertainty inequality   
\begin{equation}
\delta x \delta t \geq \alpha^\prime.  
\end{equation} 
Although non-commutators $\theta^{i j}$  are proportional to the string
tension constant $\alpha^\prime$,  they depend on velocity too. This is
quite different from the ordinary quantum mechanics  except  there
were reason we could rescale coordinates in terms of $b_1$. It might be
more  questionary that the separation $r$ is not in 5-direction but
essentially in the 9- direction. We expect that there will appear  more
convincing interpretation. 

{\sl Acknowledgment:}The work is supported in part by the National
Nature Science fund of China.

\end{document}